\documentclass[letter]{aa} 

\usepackage{graphicx}
\usepackage{natbib}
\usepackage{color}
\usepackage{upgreek}
\usepackage{soul}

\bibpunct{(}{)}{;}{a}{}{,}

\usepackage{txfonts}
\usepackage{mathrsfs}
\usepackage[colorlinks=true, linkcolor=blue, citecolor=blue, filecolor=blue, urlcolor=blue]{hyperref}
\begin{document}

\titlerunning{Magnetic fields of Be stars}
\authorrunning{S.~Hubrig et al.}

\title{The magnetic fields in Be stars are stronger than previously suggested}

\author{
S.~Hubrig\inst{1}
          \and
          S.~P.~J\"arvinen\inst{1}
          \and
          M.~Sch\"oller\inst{2}
          \and
          I.~Ilyin\inst{1}
                              }
\institute{
Leibniz-Institut f\"ur Astrophysik Potsdam (AIP),
An der Sternwarte~16, 14482~Potsdam, Germany\\
 \email{shubrig@aip.de}
              \and
European Southern Observatory, Karl-Schwarzschild-Str.~2, 85748~Garching, Germany\\            
 }

 
  \abstract
  {
Recent observational studies suggest that Be stars most likely are formed through the process of mass transfer in binary systems.
In view of the wide consensus that the origin of the magnetic field in stars with radiative envelopes
involves binary interaction processes,
searching for magnetic fields in Be stars appears especially promising.
  }
   {
     As a pilot project, we searched for the presence of magnetic fields in a sample of seven well-known Be stars.
     }
{
  We used high-resolution HARPS\-pol spectra to measure the mean longitudinal magnetic field,
  employing the least squares deconvolution technique.
A dedicated measurement procedure introduced by our group in recent years was applied.}
{
Opposite to previous spectropolarimetric studies reporting that magnetic fields in Be stars are weak and usually below 100\,G,
our study presents the first observational evidence that magnetic fields in Be stars can be as strong as
a few hundred gauss.
Magnetic fields are detected in all studied Be stars, with the strongest magnetic field being about $-$460\,G
for the B0.5\,III star HD\,184915.
Magnetic fields in the range between 338 and 380\,G (in absolute values) are detected in three other Be stars,
HD\,209409, HD\,209522, and HD\,224686.
Due to the fact that magnetic fields in Be stars are stronger than previously believed, we must re-evaluate our understanding of the
initial conditions of massive binaries to be able to determine the origin of such systems.}
   {}

\maketitle
%

   \section{Introduction}
   \label{sect:intro}

Be stars are defined as rapidly rotating main sequence stars showing normal
   B-type spectra with superposed Balmer line emission. These stars are furthermore
characterized by the episodic dissipation and formation of a
        circumstellar disk-like environment called the Keplerian decretion disk, non-radial
pulsations (NRPs), and photometric and spectroscopic variability.
Several scenarios have been suggested to explain the accumulation of material in equatorial
disks around Be stars. \cite{Cassinelli2002} suggested a magnetically torqued disk model, in which a
sufficiently strong magnetic field (of the order of 300\,G) channels a flow of wind material towards the
equatorial plane to form a disk. The study of a variety of stellar models by \cite{Maheswaran2009}
showed that in the disk region that is initially formed when wind material is channeled by dipole-type magnetic
fields towards the equatorial plane, magnetorotational instability  can set in and assist in the formation of
a quasi-steady disk. Magnetic fields as weak as a few tens of gauss will be able to channel
wind flow into a proto-disk region.
Another scenario involves NRPs, which seem to be a key aspect of the mass
ejection mechanism (e.g. \citealt{Labadie2022}; \citealt{Saio2024}).
On the other hand, numerous, more recent studies indicate that Be stars most likely form
through mass transfer in binary systems, and many of them are now seen as parts of binary
systems comprising a lower-mass stripped-envelope companion star (e.g. \citealt{Labadie2025}).
In this case, it is possible that the rapid rotation of Be stars is related to the
binarity interaction and that their decretion disk is formed due to the impact of accretion.
Yet, according to \cite{Xing2026}, mass transfer between two non-degenerate stars is an
insufficiently understood process within the framework of binary evolution, and a detailed description
of the accretion process would require dedicated multi-dimensional hydrodynamical simulations.

Observational evidence that a significant fraction of Be stars are products of binary mass transfer
is especially intriguing in view of the wide consensus that the origin of the magnetic
field in stars with radiative envelopes involves mass transfer, a common
envelope evolution, or a merging event (e.g. \citealt{Tout2008}; \citealt{Ferrario2009}; \citealt{Schneider2016}).
On the observational side, \cite{Hubrig2023,Hubrig2026} analyzed representative samples of O- and B-type
systems at different stages of interaction using the HARPS\-pol spectrograph
attached to the ESO 3.6\,m telescope \citep{Harps2008} and, based on the
numerous magnetic field detections,
concluded that interaction between the system components plays a definite role in the generation of magnetic
fields in massive O and B stars.
Previous analyses of low-resolution spectropolarimetric data of Be stars obtained with the ESO multi-mode
instrument FORS1 by \cite{Hubrig2007,Hubrig2009} revealed the presence of weak magnetic fields
in seven Be stars: HD\,62367, $\mu$\,Cen, $o$\,Aqr, $\epsilon$\,Tuc, 27\,CMa, $\chi$\,Oph, and V\,1075\,Sco. These studies suggested that
the magnetic fields of Be stars are not strong, usually below 100\,G. The strongest longitudinal magnetic field, $\left< B_{\rm z} \right>=146\pm32$\,G,
was detected in the Be star 27\,CMa.
Later on, \cite{Wade2016} analyzed high-resolution spectropolarimetric data for 85 Be stars
using the multi-line least-squares deconvolution (LSD) method  \citep{Donati1997}
and found no evidence of a magnetic field in any of the studied stars. 
Notably, \cite{Wade2016}  used masks with photospheric lines belonging to different elements
altogether, while the more recent analyses by \cite{Hubrig2023,Hubrig2026}
followed a dedicated procedure involving
for each target different line masks populated for each element separately.

To test the influence of interaction processes on the generation of
magnetic fields in massive stars,
we initiated as a pilot project a spectropolarimetric study of
a sample of seven well-known Be stars. 
The paper is laid out as follows: We
 present our HARPS\-pol spectropolarimetric observations, their reduction,
 and the magnetic field measurements for each individual target in Sect.~\ref{sect:mfield}.
 The results of the measurements of the studied Be stars, the distribution of their magnetic field strengths,
and the implication of the field detections for the origin of Be stars are briefly discussed in
 Sect.~\ref{sect:disc}.

  \begin{table*}
\caption{
Observations and results of the magnetic field
measurements for all targets in our sample. 
}
\label{tab:obsall}
\centering
\begin{tabular}{rccccrccc r@{$\pm$}l}
\hline\hline \noalign{\smallskip}
\multicolumn{1}{c}{HD} &
\multicolumn{1}{c}{Spectral} &
\multicolumn{1}{c}{$m_{\rm V}$} &
\multicolumn{1}{c}{MJD} &
\multicolumn{1}{c}{Instr.} &
\multicolumn{1}{c}{$S/N$} &
\multicolumn{1}{c}{Line} &
\multicolumn{1}{c}{FAP}&
\multicolumn{1}{c}{Det.} &
\multicolumn{2}{c}{$\left< B \right>_{\rm z}$} \\
\multicolumn{1}{c}{number} &
\multicolumn{1}{c}{type} &
\multicolumn{1}{c}{} &
\multicolumn{1}{c}{} &
\multicolumn{1}{c}{} &
\multicolumn{1}{c}{} &
\multicolumn{1}{c}{mask} &
\multicolumn{1}{c}{} &
\multicolumn{1}{c}{flag} &
\multicolumn{2}{c}{[G]} \\
\noalign{\smallskip}\hline \noalign{\smallskip}
14850  & B8IIIe         & 8.40 &  60865.418209 & H & 162  & HeSiFe     & $9\times10^{-7}$ & DD & $-$59  & 84   \\
184915 & B0.5IIIn       & 4.97 &  60862.387233 & H & 245  & HeMgSi     & $1\times10^{-6}$ & DD & $-$461 & 59   \\
205637 & B3V            & 4.55 &  55723.535672 & E & 481  & HeSiFe     & $<10^{-10}$      & DD & $-$85  & 11   \\
       &                &      &  60863.387768 & H & 422  & HeMgFe     & $7\times10^{-7}$ & DD & $-$28  & 7    \\
209014 & B8IIIsh+B8.5IV & 5.42 &  60863.438702 & H & 477  & MgSiFe     & $7\times10^{-6}$ & DD & 38     & 87   \\
209409 & B5V            & 4.69 &  60863.371608 & H & 413  & HeMgSiCrFe & $6\times10^{-6}$ & DD & 338    & 49   \\
209522 & B4IVe          & 5.95 &  60865.373670 & H & 332  & HeCSiFe    & $2\times10^{-6}$ & DD & $-$380 & 71   \\
224686 & B8V            & 4.47 &  60863.406779 & H & 482  & SiFe       & $6\times10^{-9}$ & DD & 375    & 109  \\
 \hline 
\end{tabular}
\tablefoot{
 \textit{Column~1}: HD number of the star. \textit{Column 2}: Spectral type taken from SIMBAD. \textit{Column 3}: Visual magnitude.
\textit{Column~4}: MJD values at the middle of the exposure. \textit{Column 5}: Instrument used, where H stands for HARPS\-pol and E for ESPaDOnS. \textit{Column 6}: Signal-to-noise ratio measured in the Stokes~$I$ spectra in the
spectral region around 5000\,\AA.
\textit{Column~7}: Applied line masks.
\textit{Column~8}: FAP values.
\textit{Column~9}: Detection flag, where DD means a definite detection.
\textit{Column~10}: Measured LSD mean longitudinal magnetic field strength.
}
\end{table*}

\section{Observations and  magnetic field measurements}
  \label{sect:mfield} 

 We used HARPS\-pol  from July~6 to July~9 in 2025 to obtain circular polarized spectra of seven Be stars.
HARPS\-pol has a resolving power of
 about 115\,000 and a wavelength coverage from
3780 to 6910\,\AA{} with a small gap between 5259 and 5337\,\AA{}.
With these data, we have access to measurements of the mean longitudinal magnetic field, $\langle B_{\rm z}\rangle$,
which is the line-of-sight component of the magnetic field weighted with the line
intensity and averaged over the visible hemisphere. 
Additionally, we retrieved from the Canada-France-Hawaii Telescope (CFHT)
science archive ESPaDOnS (the Echelle SpectroPolarimetric Device for Observations of Stars)
observations of HD\,205637 from 2011 covering the wavelength range from
3750 to 10\,500\,\AA{}  with a spectral resolution of about 65\,000.
As in our previous studies that used HARPS\-pol data (e.g.\ \citealt{Hubrig2018}; \citealt{jarvinen2018}),
we employed the LSD technique following the description given by \citet{Donati1997}
in order to increase the accuracy of the mean longitudinal magnetic field determination.
The parameters of the lines used to calculate the LSD profiles were taken
from the Vienna Atomic Line Database \citep[VALD3;][]{Kupka2011}.
All of our targets show the presence of a proper motion anomaly \citep{Kervella2019},
indicating probable membership in binary systems, as expected for Be stars.
Thus, we applied for the treatment of their spectra a dedicated procedure similar to that
described by \citet{Hubrig2023,Hubrig2026}.
Its measurements involve different line masks populated for each element separately.
Such a strategy is also frequently employed in the studies of B-type stars
hosting magnetic fields,
where abundance anomalies are associated with the presence of surface chemical patches.
The distribution of such patches is non-symmetric with respect to the rotation axis
but shows a correlation with
the magnetic field topology (e.g.\ \citealt{Rice1997,Hubrig2012,Hubrig2017}).
As the visibility of the patches is changing during the stellar rotation,
different line masks should be tested. More details about our LSD analysis are presented in
Appendix~\ref{sect:app1}.

The presence of a magnetic field in the LSD profile was evaluated according to \citet{Donati1992},
who defined a Zeeman profile with a false alarm probability (FAP) of $\leq 10^{-5}$ as a definite detection,
whereas $10^{-5} <$ FAP $\leq 10^{-3}$ is a marginal detection, and FAP $> 10^{-3}$ is a non-detection.
A summary of our measurements is presented in Table~\ref{tab:obsall}.
In the following, we discuss for each target the spectral appearance, previously
known properties, and the results of the measurements.

\paragraph*{HD\,14850 (=CPD$-$30\,286):}

\begin{figure}
    \centering
    \includegraphics[width=0.24\textwidth]{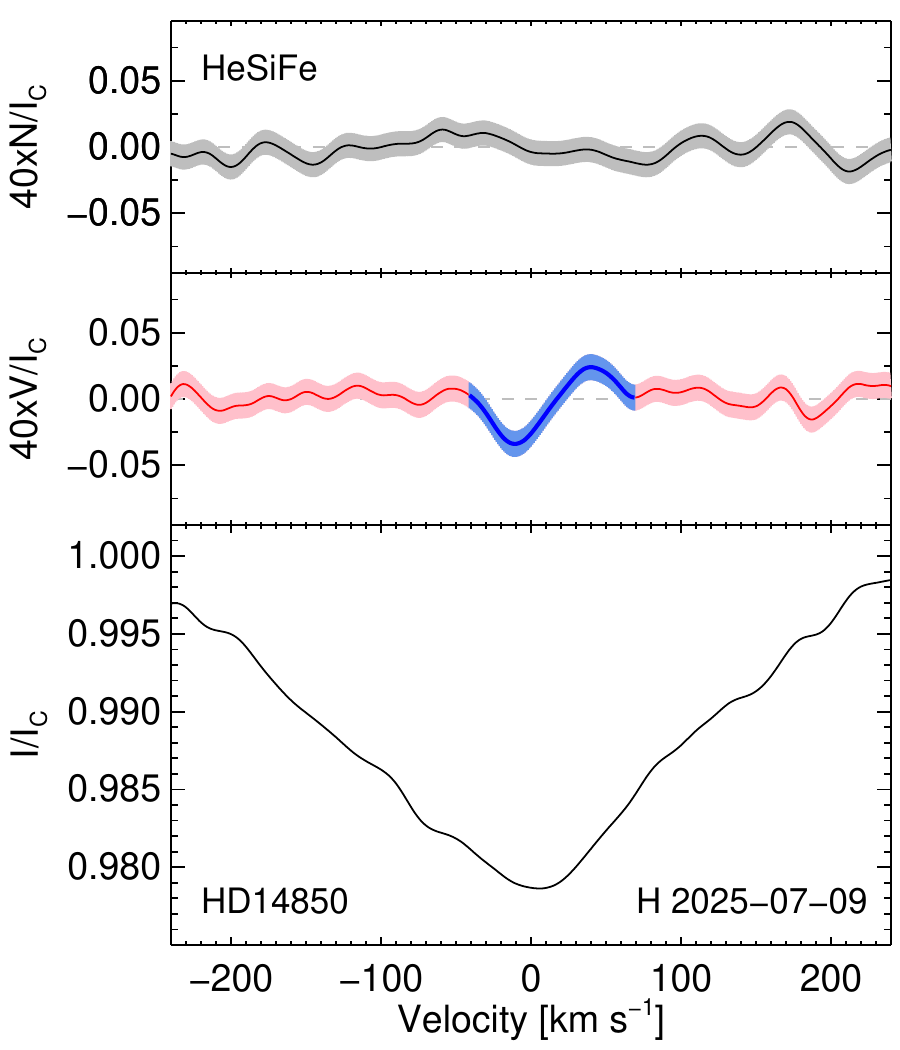}
 \includegraphics[width=0.24\textwidth]{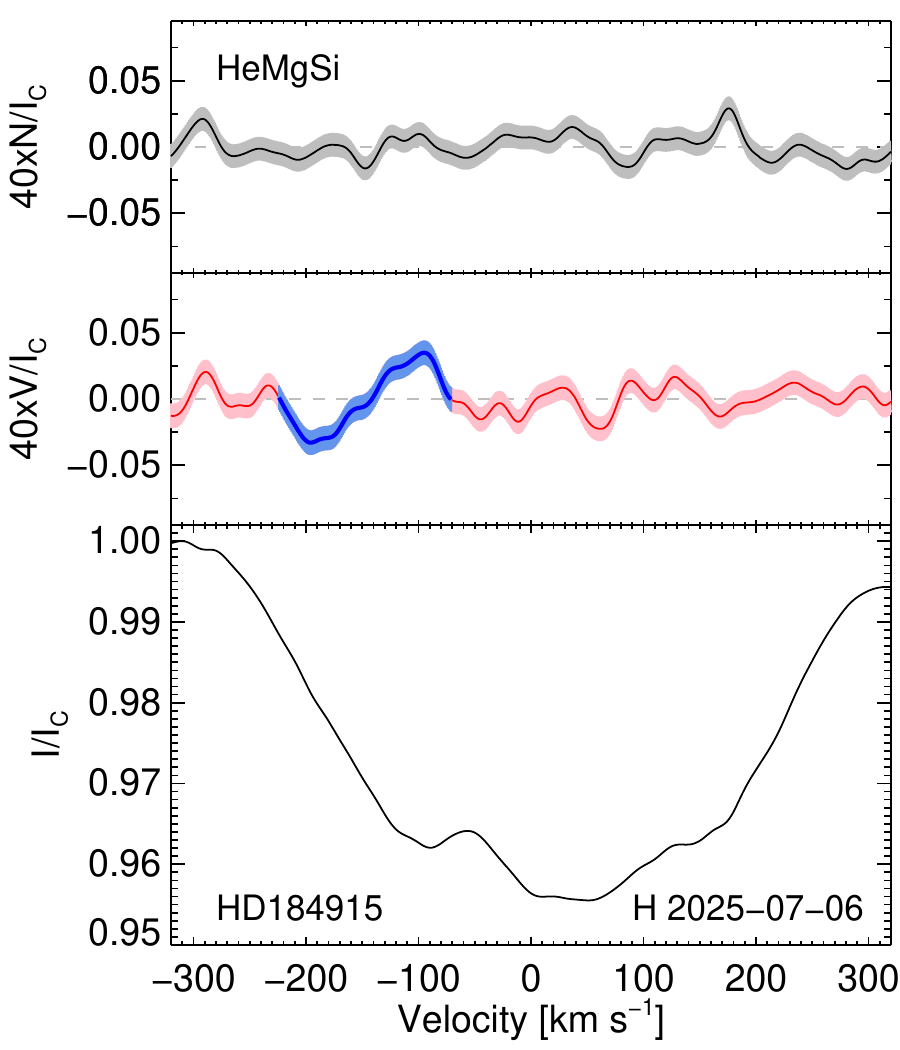}
    \caption{Results of the LSD analysis for the Be stars HD\,14850 (\textit{left}) and HD\,184915 (\textit{right}). Stokes~$I$, Stokes~$V$, and
      diagnostic null $N$ spectra were obtained for
      the line masks indicated in the plots. The identified Zeeman signatures are in blue.
   }
    \label{fig:IVN14850}
\end{figure}

The spectrum of this B8\,IIIe high Galactic latitude star \citep{Slettebak1997}, shown in Fig.~\ref{fig:reg14850},
exhibits strong emission  in the H$\alpha$ line and emission in the \ion{Fe}{ii} lines at 6456 and 6516\,\AA{}, which are frequently identified in the spectra of Be stars.
We achieved for this star a definite
magnetic field detection with $\left< B_{\rm z} \right>=-59\pm84$\,G using a mask containing He, Si, and Fe lines. Our LSD analysis
is presented in Table~\ref{tab:obsall} and the left panel of Fig.~\ref{fig:IVN14850}.

\paragraph*{HD\,184915 (=$\kappa$\,Aql):}

As we show in Fig.~\ref{fig:reg184915}, we detect in our HARPS\-pol spectrum of this target weak emission on both sides of the H$\alpha$ line
and in the blue wings of the \ion{Si}{iii} triplet lines.
Notably, numerous lines show a distinct
structure, which can probably be explained by the presence of a companion
or the presence of NRPs similar to pulsations observed
in $\beta$~Cep and slowly pulsating B stars.
The presence of NRPs is frequently determined through observations of moving peaks
in the cores of spectral lines. From the theoretical considerations presented by \cite{Hubrig2004}, it
follows that pulsationally modulated variations of
the order of 100\,G may exist in the outer atmospheric layers of stars with kilogauss-scale magnetic fields.
On the observational side, \cite{Shultz2017} studied the impact of pulsations for the $\beta$\,Cephei pulsator HD\,46328
and showed that at some observing epochs the measured longitudinal magnetic field can vary by about  $\pm2-3\%$.

According to \cite{Burgos2024}, this target is enriched in helium, suggesting that it might be a product
of binary evolution.
Of all the targets studied, HD\,184915 possesses the strongest longitudinal magnetic field,
$\left< B_{\rm z} \right>= -461\pm59$\,G. The field was definitely detected using a mask containing He, Mg, and Si lines.
Our LSD analysis is presented in Table~\ref{tab:obsall} and the right panel of Fig.~\ref{fig:IVN14850}.

\paragraph*{HD\,205637 (=$\epsilon$\,Cap):}

  \begin{figure}
    \centering
    \includegraphics[width=0.240\textwidth]{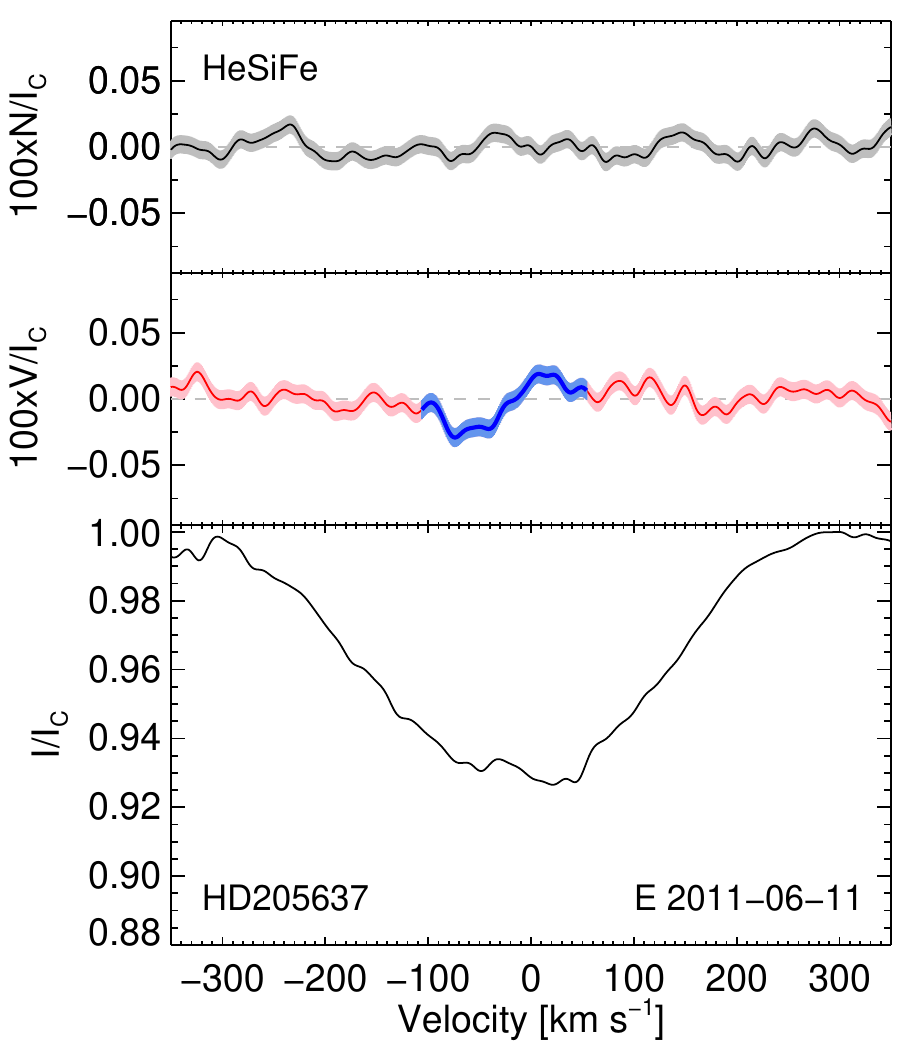}
    \includegraphics[width=0.240\textwidth]{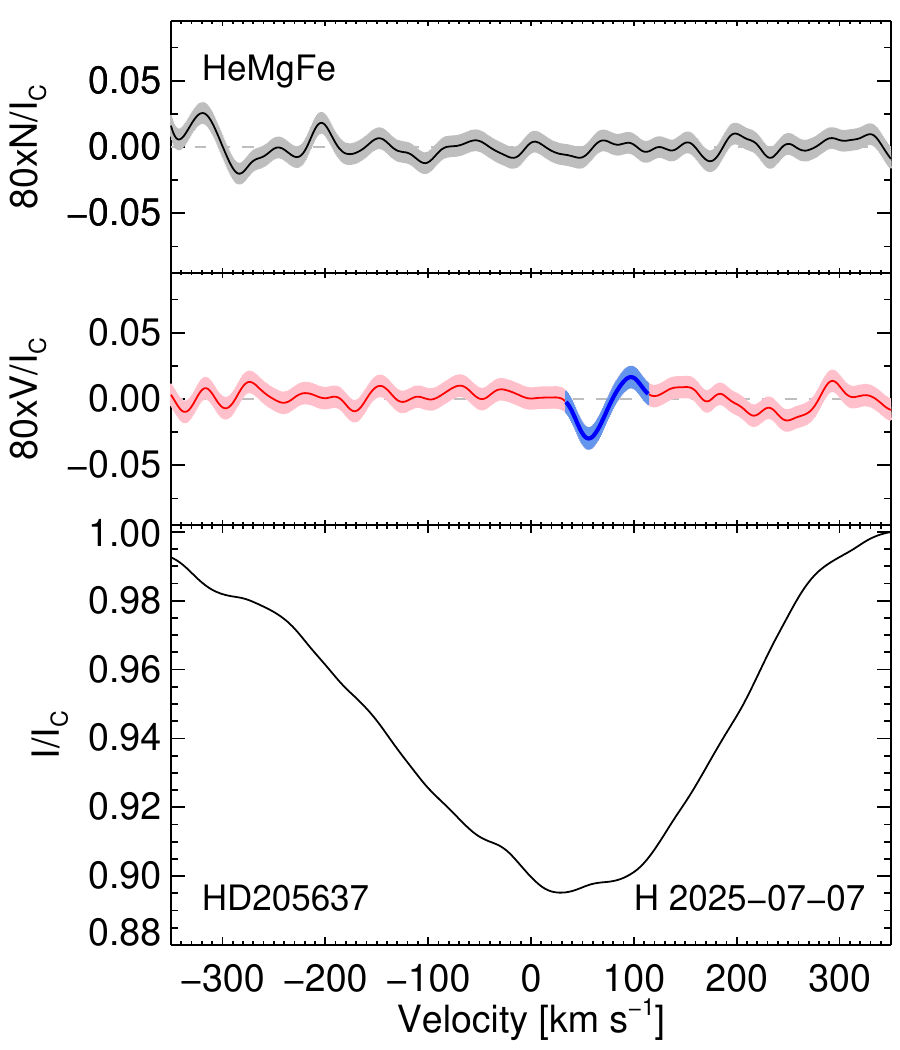}
    \caption{LSD analysis of HD\,205637 using ESPaDOnS (\textit{left}) and HARPS\-pol (\textit{right}) observations.
    }
    \label{fig:IVN205637}
\end{figure}

The presence of a companion with an orbital period of 128.5\,d was previously reported in  \cite{Rivinius2006}.
Our spectra also show a change in the radial velocity between 2011 and 2025.
Apart from the HARPS\-pol observation, this target was also observed on one night in June 2011 with ESPaDOnS.
It was recorded five times using consecutive short exposures with 10\,min durations.
Figure~\ref{fig:reg205637} shows that the emission in the H$\alpha$
and \ion{He}{i}~6678 line profiles were stronger in 2011 than in 2025.
It is of interest that \cite{Buscombe1969} classified this target as a shell star with sharp absorption lines formed in the surrounding gaseous shell.  It has been
 suggested that the appearance of a Be star as a shell star can be determined
from the inclination angle of the disk (e.g.\ \citealt{Hanuschik1995}).
No changes between the five ESPaDOnS spectra acquired over 0.8\,h have been detected.
We achieved for HD\,205637 definite detections with $\left< B_{\rm z} \right>=-85\pm11$\,G using the ESPaDOnS observations with a He, Si, and Fe
mask and $\left< B_{\rm z} \right>=-28\pm7$\,G using the HARPS\-pol observations and a mask containing He, Mg, and Fe lines.
The results of our analysis are presented in Table~\ref{tab:obsall} and Fig.~\ref{fig:IVN205637}.

\paragraph*{HD\,209014 (=$\eta$\,PsA):}

\begin{figure}
    \centering
    \includegraphics[width=0.24\textwidth]{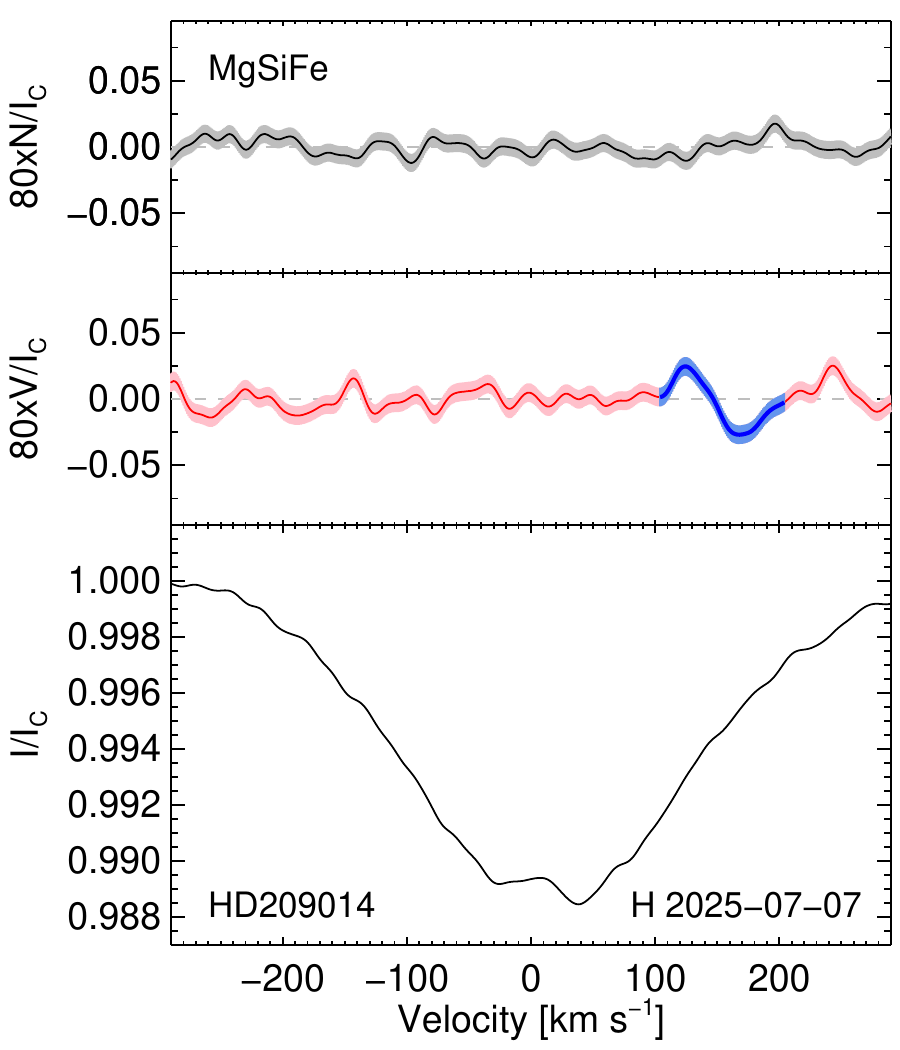}
 \includegraphics[width=0.24\textwidth]{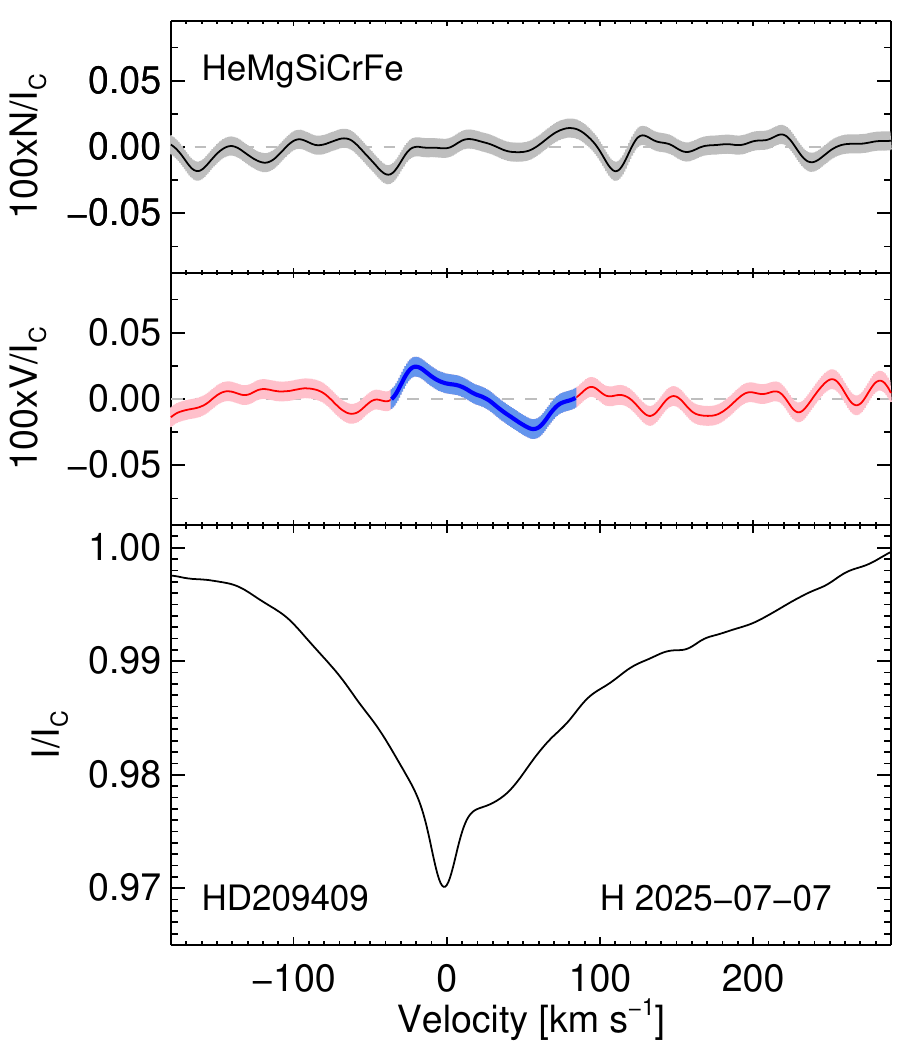}
    \caption{Same as Fig.~\ref{fig:IVN14850} but for HD\,209014 (\textit{left}) and HD\,209409 (\textit{right}).
}
    \label{fig:IVN209014}
\end{figure}

According to \citet{Corbally1984}, this target is a close visual binary system
with a primary B8\,IIIsh star and a B8.5\,IV companion.
As we show in Fig.~\ref{fig:reg209014}, the H$\alpha$ line in the HARPS\-pol spectrum is in emission and the profile of the \ion{Fe}{ii}~5018 line
displays emission in both the red and blue wings. 
We achieved for this target a definite magnetic field detection with $\left< B_{\rm z} \right>=38\pm87$\,G using a mask containing Mg, Si, and Fe lines.
The results of our analysis are presented in Table~\ref{tab:obsall}
and the left panel of Fig.~\ref{fig:IVN209014}.

\paragraph*{HD\,209409 (=$o$\,Aqr):}

 \cite{Cowley2015} consider this bright target to be a prototype Be shell star.
While H$\alpha$ is in emission and the \ion{Fe}{ii}~5018 profile
in Fig.~\ref{fig:reg209409} shows a typical sharp shell core 
together with emission wings, other Balmer lines exhibit sharp absorption cores,
much sharper than expected from the widths of normal photospheric lines.
The \ion{He}{i} lines are broad and shallow, unlike the metal lines, indicating an atmospheric origin.
The Stokes~$I$ profile in Fig.~\ref{fig:IVN209014} shows a composite line profile structure. 
Using low-resolution spectropolarimetric observations with FORS1,
\cite{Hubrig2009} detected a weak  magnetic field,
$\left< B_{\rm z} \right>=-98\pm31$\,G. 
Our analysis of high-resolution HARPS\-pol observations confirms the magnetic nature
of HD\,209409 thanks to our definite
magnetic field detection with $\left< B_{\rm z} \right>=338\pm49$\,G
using a mask containing He, Mg, Si, Cr, and Fe lines.
The results of our analysis are presented in Table~\ref{tab:obsall}
and the right panel of Fig.~\ref{fig:IVN209014}.

\paragraph*{HD\,209522 (=UU\,PsA):}

\begin{figure}
    \centering
    \includegraphics[width=0.24\textwidth]{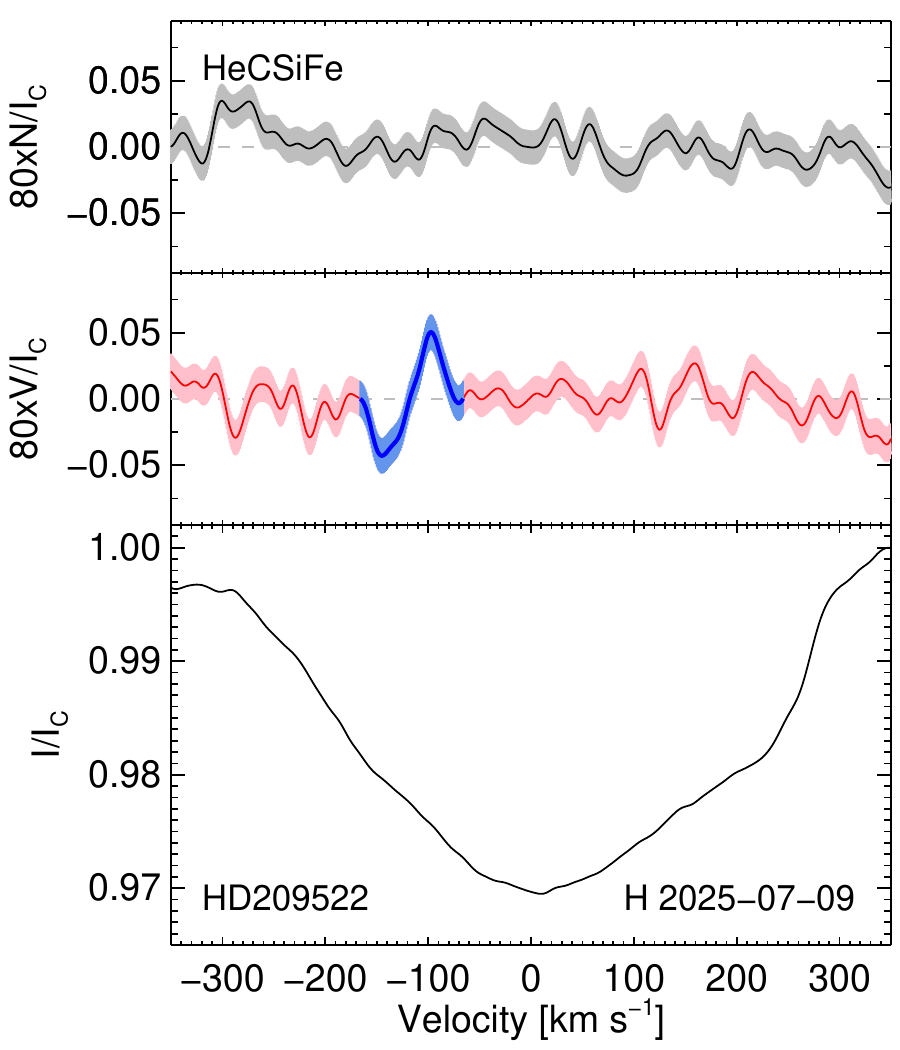}
 \includegraphics[width=0.24\textwidth]{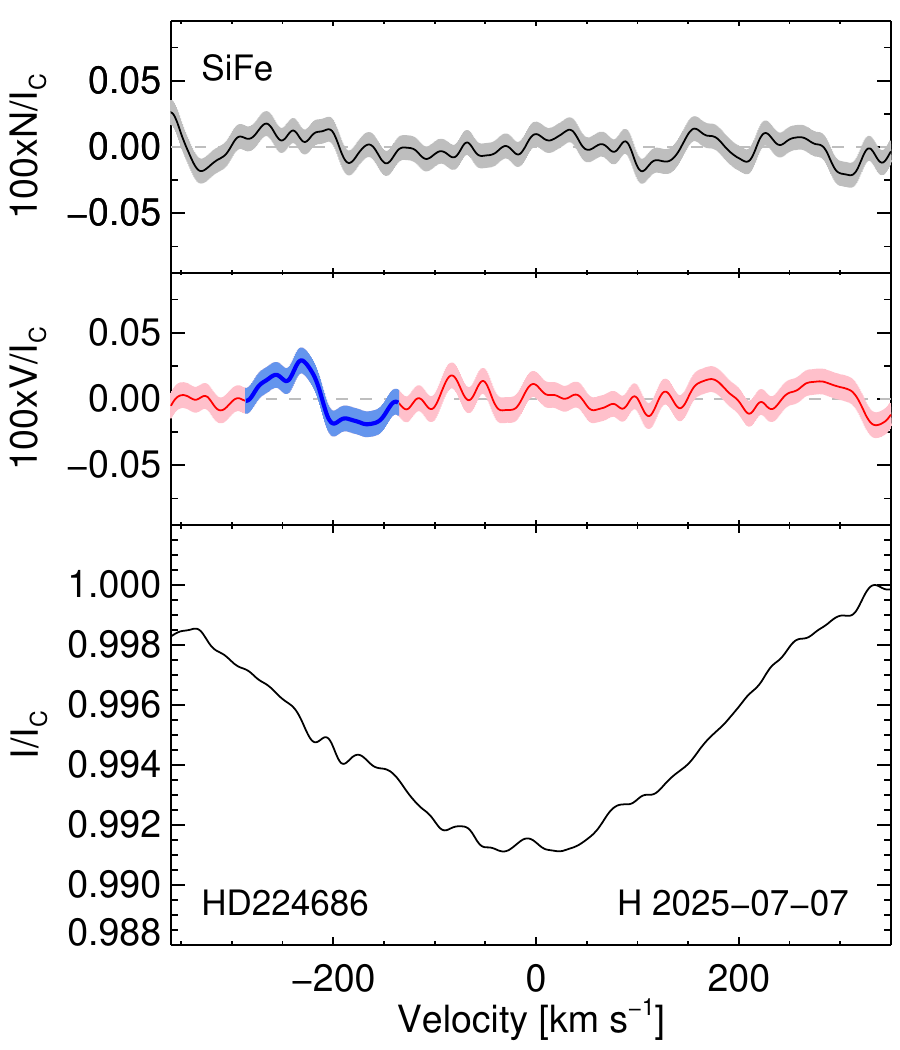}
    \caption{Same as Fig.~\ref{fig:IVN14850} but for HD\,209522 (\textit{left}) and HD\,224686 (\textit{right}).
}
    \label{fig:IVN209522}
\end{figure}

According to \citet{Shokry2018}, who studied the stellar parameters of Be stars observed with X-shooter, no recent study has found Balmer emission in this star.
However, H$\alpha$ emission is clearly detected in our HARPS\-pol spectrum (see Fig.~\ref{fig:reg209522}). We also observe
a weak emission in the red wing of the \ion{Fe}{ii} line at 5018\,\AA{}. We achieved for this star a definite
magnetic field detection with $\left< B_{\rm z} \right>=-380\pm71$\,G using a mask containing He, C, Si, and Fe lines.
The results of our analysis are presented in Table~\ref{tab:obsall} and the left panel of Fig.~\ref{fig:IVN209522}.

\paragraph*{HD\,224686 (=$\epsilon$\,Tuc):}

As shown in Fig.~\ref{fig:reg224686}, the HARPS\-pol spectrum of this target
displays characteristics of shell stars, with the H$\alpha$ line 
exhibiting sharp absorption cores, much sharper than expected from the width of normal photospheric lines.
Furthermore, \ion{Fe}{ii} lines belonging to Multiplet~42 show the presence of central emission.
While \cite{Buscombe1969} classified this target as B7Vke with a note that H$\alpha$ is in emission,
almost a decade later,
\cite{Pedersen1978} presented an H$\alpha$ line profile similar to that detected in our observations.
The detection of a weak longitudinal magnetic field in HD\,224686 with $\left< B_{\rm z} \right>=74\pm24$\,G
was previously reported by \cite{Hubrig2009}.
We achieved for this star a definite magnetic field detection with $\left< B_{\rm z} \right>=375\pm109$\,G
using a mask containing Si and Fe lines.
The results of our analysis are presented in Table~\ref{tab:obsall} and the right panel of
Fig.~\ref{fig:IVN209522}.

\section{Discussion}
\label{sect:disc}

In this work, we present observational evidence that Be stars can possess magnetic fields of the order of a
few hundred gauss.
We also confirm the presence of a magnetic field in two targets,
HD\,209409 and HD\,224686, previously observed with FORS1.
Magnetic fields in five Be stars have been measured for the first time. HD\,184915 has the strongest magnetic field, of about $-$460\,G, 
followed by HD\,209409, HD\,209522, and HD\,224686 with field strengths  above 300\,G in absolute values.
Apart from HD\,205637, our targets have been observed only once. Since, per definition, the longitudinal magnetic field is strongly dependent on the
viewing angle between the field orientation and the observer and is modulated as the star rotates,
its magnetic field could be even stronger.

In all LSD plots, the detected Zeeman
signatures do not extend over the full LSD Stokes~$I$ profile. As mentioned in Sect.~\ref{sect:mfield}, this is explained by the fact
that magnetic A- and B-type stars usually show a rotationally modulated appearance of chemical patches.
On the other hand, since all targets in our sample
are reported to show a proper motion anomaly, the shifted locations of the observed
Zeeman signatures relative to the underlying Stokes~$I$ profiles probably indicate that some of our targets are close binaries with magnetic components.

Four  targets in our sample, HD\,205637, HD\,209014, HD\,209409, and HD\,224686, currently display or displayed characteristics of shell stars.
Importantly, magnetic fields have previously been detected in two other shell Be stars, the visual binary
27\,CMa with  $\left< B_{\rm z} \right>=-146\pm32$\,G
using FORS1 spectropolarimetry \citep{Hubrig2007} and in HD\,61954 using HARPS\-pol observations  with $\left< B_{\rm z} \right>=156\pm46$\,G  and
$\left< B_{\rm z} \right>=-420\pm95$\,G measured on two different nights \citep{Hubrig2025}.
HD\,61954 is, however, a well-known blue straggler star in the open cluster NGC\,2437, for which  mass transfer from a binary companion has probably led to the
rejuvenation of the mass gainer.
Based on the reported magnetic field detections in targets with shell spectra,
it is possible that other Be stars with shell-like line profiles
are also products of interaction and therefore possess magnetic fields.

The origin of fast-rotating Be stars with decretion disks remains unclear.
According to \cite{Labadie2025}, there are two classes of Be binaries,
pre-interaction and post-interaction.
Given the detection of magnetic fields in Be stars,
a crucial improvement of our understanding of the initial conditions of massive binaries
is required to be able to determine the origin of such systems.

\begin{acknowledgements}

This work is based on observations made with ESO telescopes at the
La Silla Paranal Observatory under programme ID 0115.D-2108(A). 
Observations of HD\,205637 were also obtained with CFHT,
which is operated by the National Research Council of Canada,
the Institut National des Sciences de l'Univers of the Centre National de la Recherche Scientifique of France,
and the University of Hawaii.

\end{acknowledgements}

%
   \bibliographystyle{aa} 
   \bibliography{aa60036-26} 
%

\begin{appendix}

\section{Selection of spectral lines to populate line masks}
   \label{sect:app1}

   Significant differences in the magnetic field strengths measured
in B-type stars using spectral lines belonging to different elements were previously
mentioned in numerous studies (e.g.\ \citealt{Hubrig2017,Shultz2018}).
As shown by these authors, magnetic field phase curves obtained using spectral lines of individual
elements can even show variability in antiphase. This is possible if chemical abundance patches 
with a concentration of individual elements have a different distribution in horizontal or vertical direction,
depending on the magnetic field configuration  and the atmospheric structure of the star.
Taking into account that lines belonging to different elements can show dispersion in the measured longitudinal
magnetic fields, we selected for the masks only the lines of the elements generating the strongest magnetic signal in the Stokes~$V$ spectra.
Furthermore, only the best lines that appear to be unblended or
minimally blended in the Stokes~$I$ spectra were included in the line mask.

  \section{Characteristic spectral lines in individual targets}
   \label{sect:app2}

\begin{figure}
    \centering
    \includegraphics[width=0.485\textwidth]{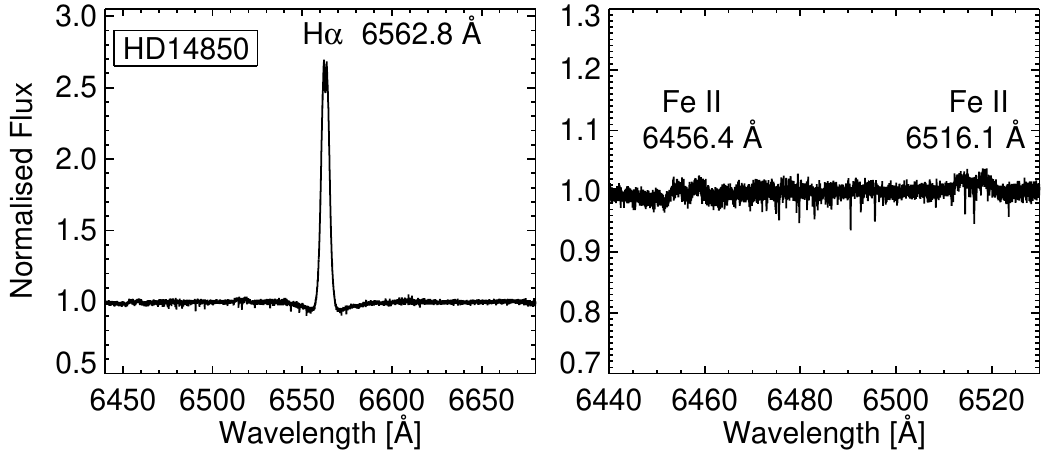}
    \caption{HARPS\-pol Stokes~$I$ spectrum of the B8\,IIIe star HD\,14850, which
exhibits emission in the H$\alpha$ line and the \ion{Fe}{ii} lines at
  6456 and 6516\,\AA{}.
These emission lines are frequently identified in the spectra of Be stars.
    }
    \label{fig:reg14850}
\end{figure}

\begin{figure}
    \centering
\includegraphics[width=0.240\textwidth]{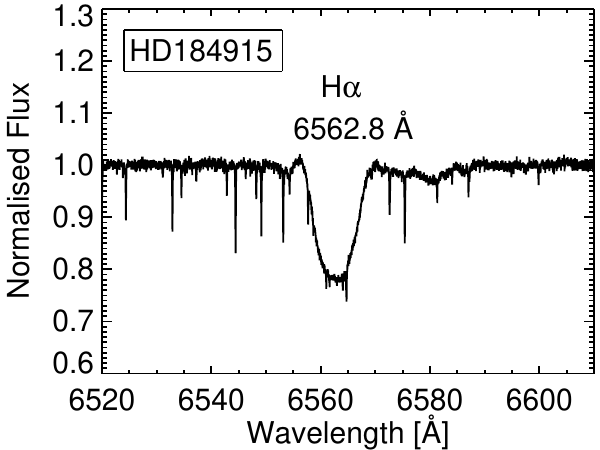}
\includegraphics[width=0.240\textwidth]{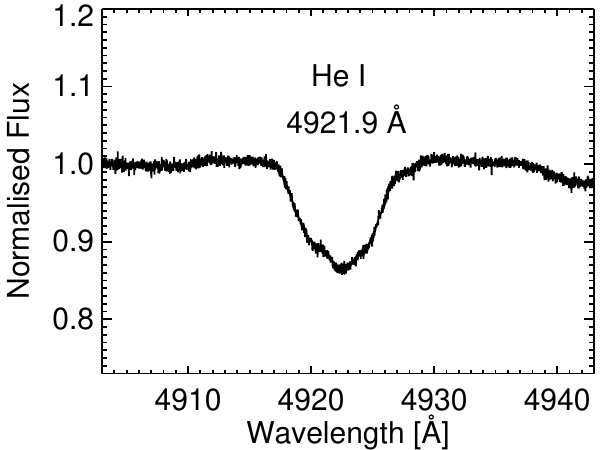}
\includegraphics[width=0.240\textwidth]{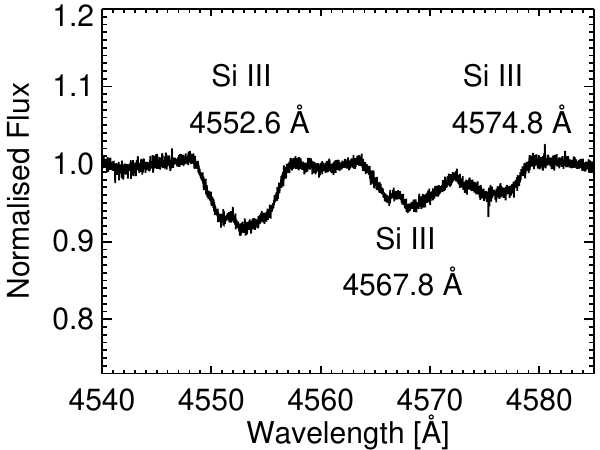}
     \caption{ HARPS\-pol Stokes~$I$ spectrum
of the B0.5\,IIIn star HD\,184915, which
shows the presence of weak emission on both sides
     of the H$\alpha$ line and in the blue wings of the \ion{Si}{iii} triplet lines.
        }
    \label{fig:reg184915}
\end{figure}

\begin{figure}
    \centering
\includegraphics[width=0.240\textwidth]{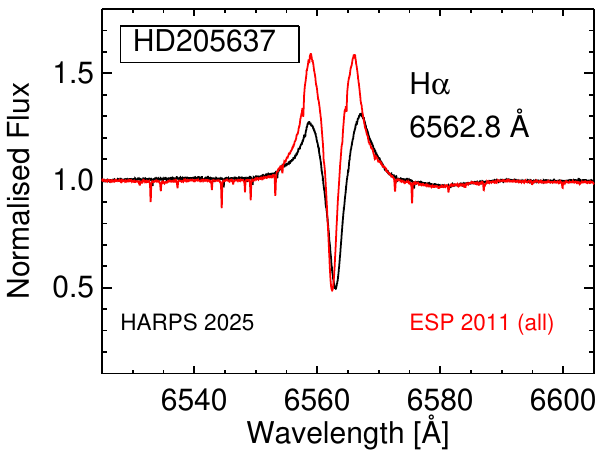}
\includegraphics[width=0.240\textwidth]{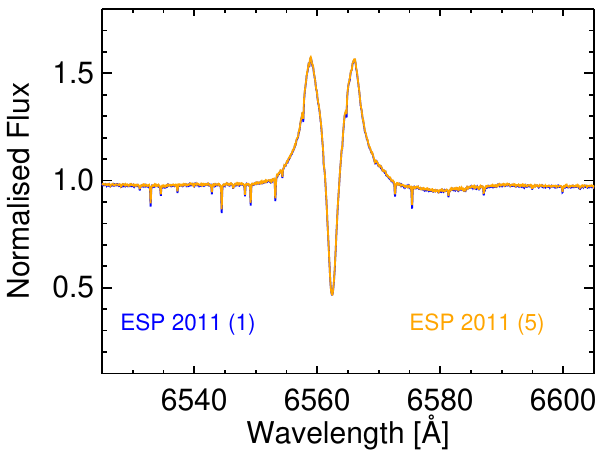}
\includegraphics[width=0.240\textwidth]{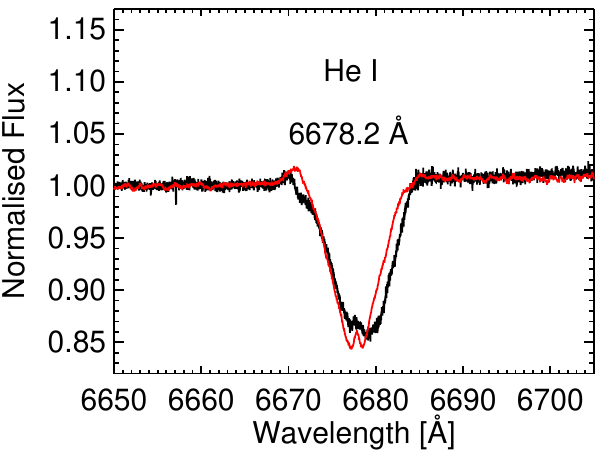}
\includegraphics[width=0.240\textwidth]{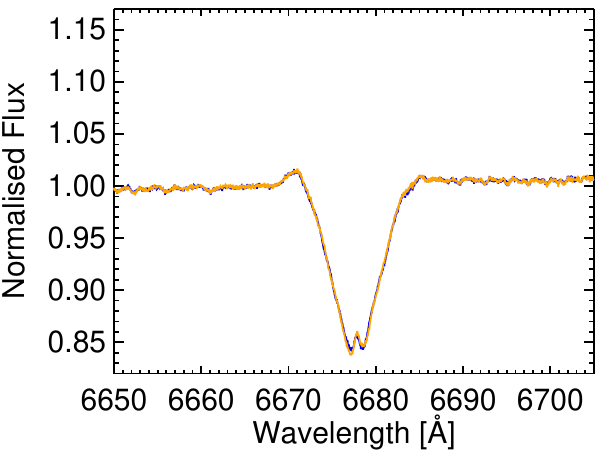}
\caption{Emission in the H$\alpha$ and the \ion{He}{i}~6678 line profiles
of HD\,205637 observed in the ESPaDOnS and HARPS\-pol Stokes~$I$ spectra.
{\it Top left:} All five ESPaDOnS H$\alpha$ profiles
overplotted with the one observed in the HARPS\-pol spectrum. 
\textit{Top right}: Only the first and last ESPaDOnS H$\alpha$ profiles are overplotted,
showing no changes over 0.8\,h.
{\it Bottom:} Same but for the \ion{He}{i}~6678 line profile.
}
    \label{fig:reg205637}
\end{figure}

\begin{figure}
    \centering
    \includegraphics[width=0.485\textwidth]{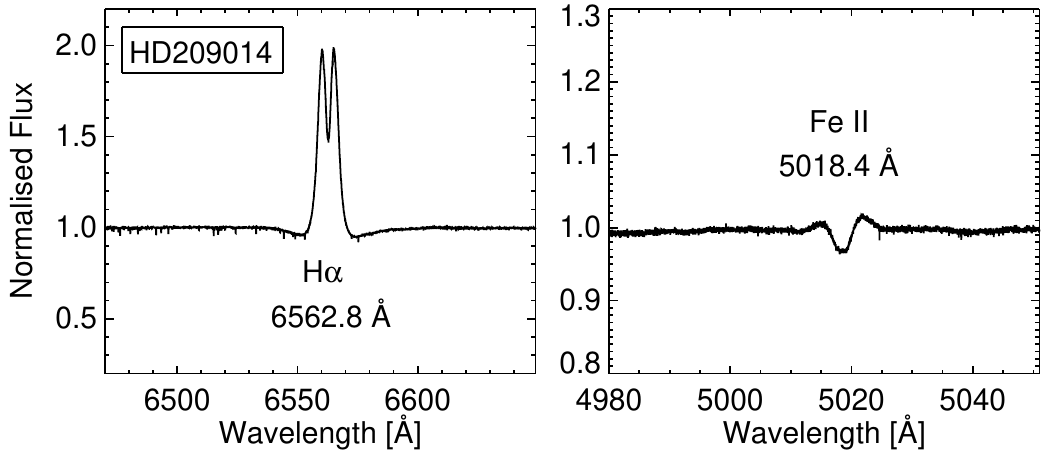}
    \caption{Emission in the H$\alpha$ line
and in both the red and blue wings
of the profile of the \ion{Fe}{ii}~5018 line
      observed in the HARPS\-pol Stokes~$I$ spectrum of HD\,209014.
    }
    \label{fig:reg209014}
\end{figure}

\begin{figure}
    \centering
    \includegraphics[width=0.485\textwidth]{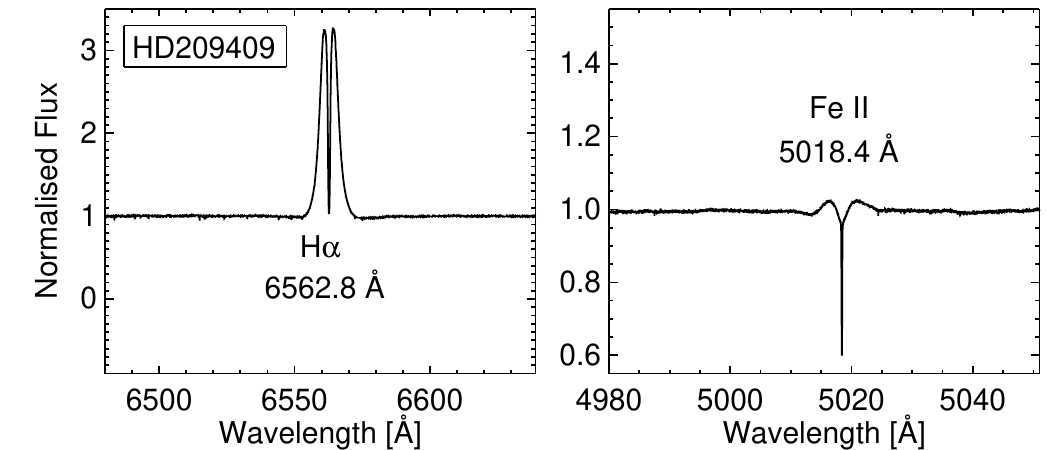}
    \caption{Strong emission in the H$\alpha$ line and the sharp core in the profile
of the \ion{Fe}{ii}~5018 line observed in the 
      HARPS\-pol Stokes~$I$ spectrum of HD\,209409.
    }
    \label{fig:reg209409}
\end{figure}

\begin{figure}
    \centering
    \includegraphics[width=0.485\textwidth]{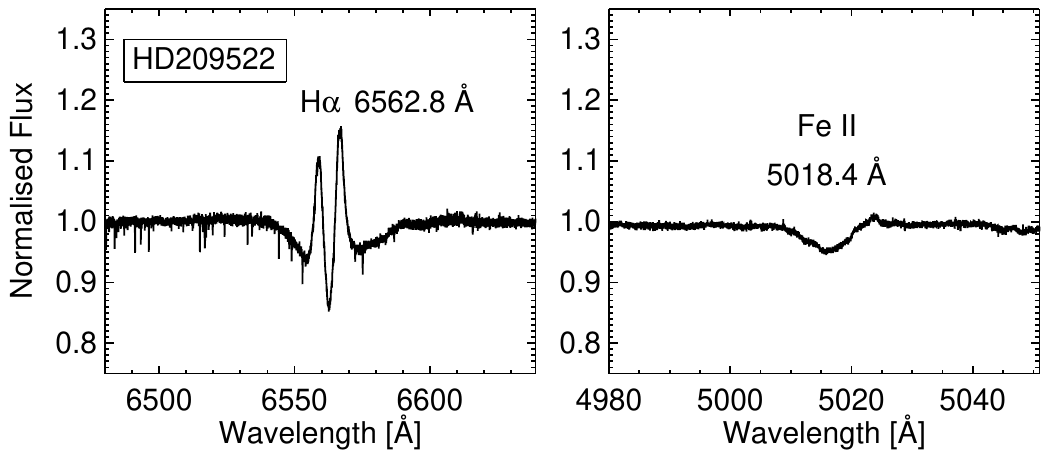}
    \caption{H$\alpha$ emission in the HARPS\-pol Stokes~$I$ spectrum of the Be star HD\,209522.
 We also observe a weak emission in the red wing of the \ion{Fe}{ii} line at 5018\,\AA{}. }
    \label{fig:reg209522}
\end{figure}

\begin{figure}
    \centering
    \includegraphics[width=0.485\textwidth]{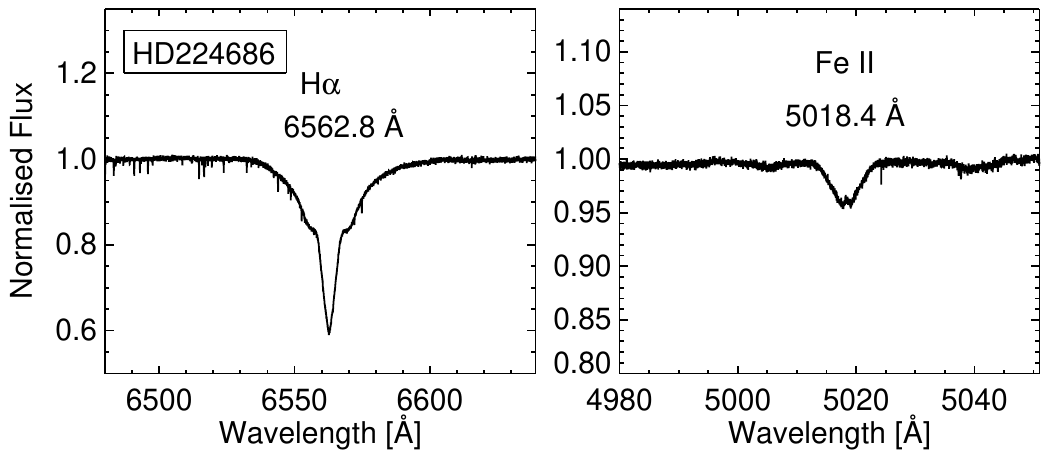}
    \caption{HARPS\-pol Stokes~$I$ spectrum of HD\,224686 displaying the typical characteristic
of a shell star, with the H$\alpha$ line exhibiting a sharp absorption core.
The profile of the \ion{Fe}{ii}~5018 line appears slightly split,
most probably due to the presence of emission.
         }
    \label{fig:reg224686}
\end{figure}

In Figs.~\ref{fig:reg14850}--\ref{fig:reg224686} we show the
integrated light (Stokes~$I$) spectra of all sources studied
in this letter.

\end{appendix}

\end{document}